\documentclass{kapproc} 

\usepackage{procps} 

\usepackage[dvips]{graphicx}

\upperandlowercase

\kluwerbib 

\newcommand{\dels}{\bar{\delta}}

\begin{document}

\articletitle[XMM-$\Omega$ project]
{XMM-$\Omega$ project : Cosmological implication from the high redshift $L - T$ relation of X-Ray clusters}

\author{Sebastien C. Vauclair and the XMM-$\Omega$ project collaboration}

\affil{Observatoire Midi-Pyrenees, 14 av. E. Belin, 31400 Toulouse, France}
\email{sebvcr@ast.obs-mip.fr}

\anxx{Vauclair\, Sebastien C.}

\begin{abstract}
The evolution of the temperature distribution function (TDF) of X-ray clusters is known to be a 
powerful cosmological test of the density parameter of 
the Universe. 
Recent XMM observations allows us to measure accurately the $L - T$ relation of high redshift X-ray clusters.
In order to investigate cosmological implication of this recent
results, we have derived theoretical number counts for different X-ray clusters samples, namely the RDCS, EMSS, SHARC, 160 deg$^2$ and MACS at $z > 0.3$ in different flat models. We show that a standard hierarchical modeling of cluster distribution in a flat low density universe , 
normalized to the local abundance, 
overproduces cluster 
abundance at high redshift ($z>0.5$) by an order of magnitude. 
We conclude that presently existing data on X-ray clusters at high redshift 
 strongly favor
a universe with a high density of matter, insensitively to the details of the modeling.
\end{abstract}


\section{Introduction}

In this work we examine the expected number counts 
with different values for the density parameter  and compare them to 
observed counts. The first model is the best flat model fitting 
 the local Temperature Distribution Function (TDF) as well as the  high redshift TDF (Henry, 1997), see Blanchard 2000. While the second model is a flat low density model normalized to the local TDF.

\section[Ingredients of the modeling]
{Ingredients of the modeling and results}
As a first step, models are normalized using the local 
temperature distribution function, 
two fundamental ingredients are needed: the mass function and 
 the mass-temperature relation, $M-T$. 
Here we use the expression of the mass function given by Sheth, Mo and Tormen,1999.
\begin{small}
\begin{equation}
f(\nu) = \sqrt{\frac{2 A}{\pi}}C\exp (-0.5A\nu^2)(1.+ 1./(A\nu)^2)^Q)
\end{equation}
\end{small} with $A = 0.707$, $C = 0.3222$, $Q = 0.3$ and $\nu\equiv\dels/\sigma(M)$.

 The $M-T$ relation is written to be:
\begin{small}
$ 
T = T_{15}(\Omega \Delta)^{1/3}M_{15}^{2/3}(1+z).
$
\end{small} \\
In this work we use different models of universe, including different $M-T$ normalizations, presented in table 1. \\
A key-ingredient of the modelling is the  $L-T$ relation and its evolution. The goal of the XMM-Omega project was to measure accurately this relation at redshift about 0.5 (Bartlett et al., 2001).
We estimate the evolution from the recent XMM observations of high redshift clusters (Lumb et al., in prepartion), following the method of Sadat et al. (1998), 
by computing for each cluster :

\begin{small}
$$
C(z) = \frac{L}{AT^B} \frac{D_l(\Omega_M=1,z)^2}{D_l(\Omega_M,z)^2}
$$
\end{small}

We parametrize the evolution by \begin{small} $C(z) = (1+z)^\beta$\end{small} and 
we determine the best fitting\begin{small} $\beta = 0.65 \pm 0.21$\end{small}, consistent with the Chandra result (Vikhlinin et al. 2002).\\
In order to compute number counts, one can notice that 
the observations actually provide $z$ and $f_x$ (rather than the actual $L_x$ and $T_x$), therefore one has to compute the following:
\begin{small}
\begin{eqnarray}
N(>f_x,z,\Delta z) = & \int_{z-\Delta z}^{z+\Delta z} \frac{\partial N}{\partial z}(L_x> 4\pi D_l^2 f_x) dz \nonumber \\
= & \int_{z-\Delta z}^{z+\Delta z} N(>T(z))dV(z) 
\end{eqnarray}
\end{small}
Results are presented in figure 1. We conclude that {\em within the standard scenario of structure formation}, the predicted abundance of galaxy clusters points toward a high density universe, insensitively to local $L - T$ used, to the dispersion on its evolution nor the different $M - T$ normalization.

\begin{figure}[!t]
\begin{center}
\begin{tabular}{c c }
\includegraphics[angle=0,totalheight=4.cm,width=5.cm]{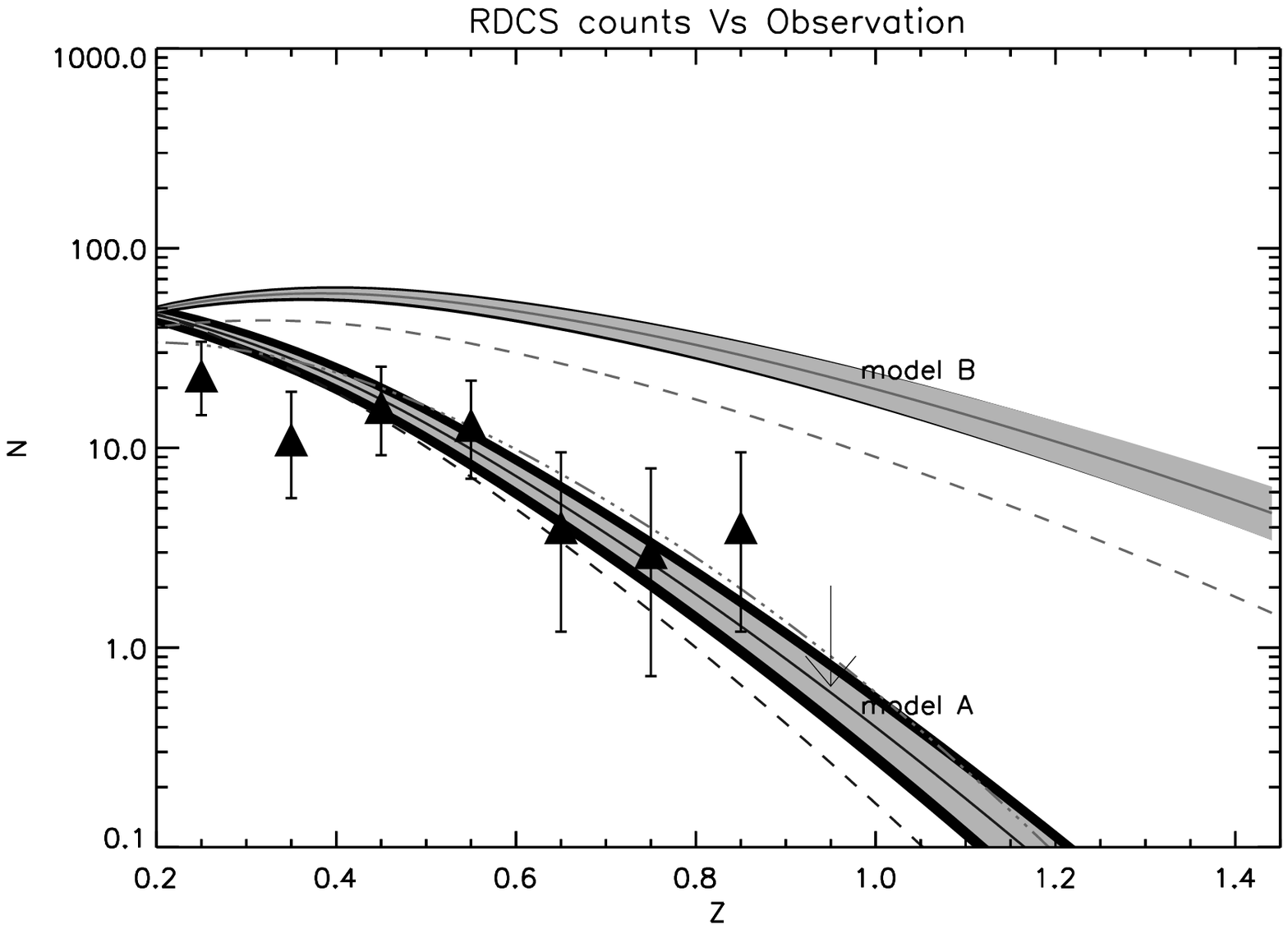}&
\includegraphics[angle=0,totalheight=4.cm,width=5.cm]{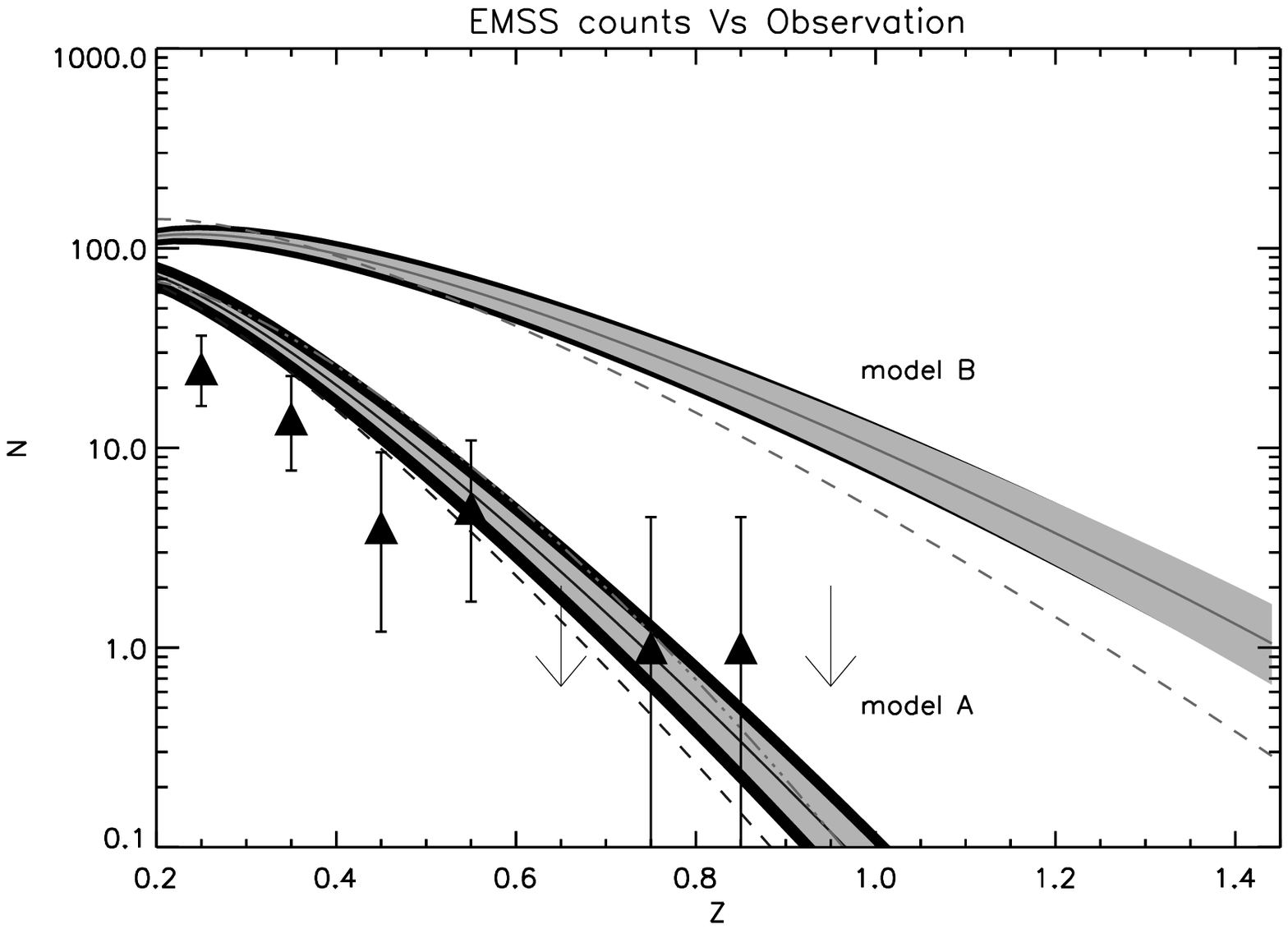}\\
\includegraphics[angle=0,totalheight=4.cm,width=5.cm]{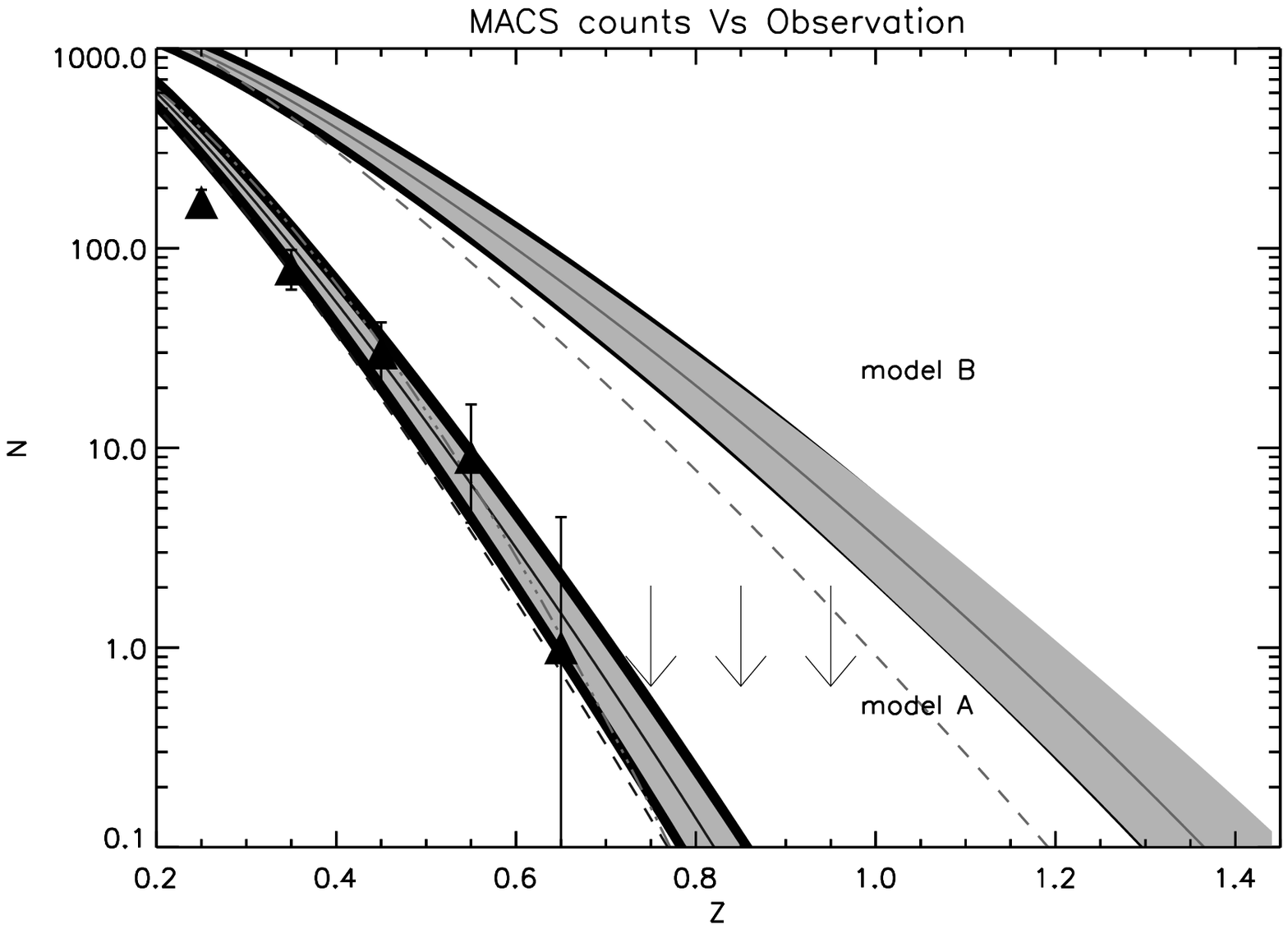}&
\includegraphics[angle=0,totalheight=4.cm,width=5.cm]{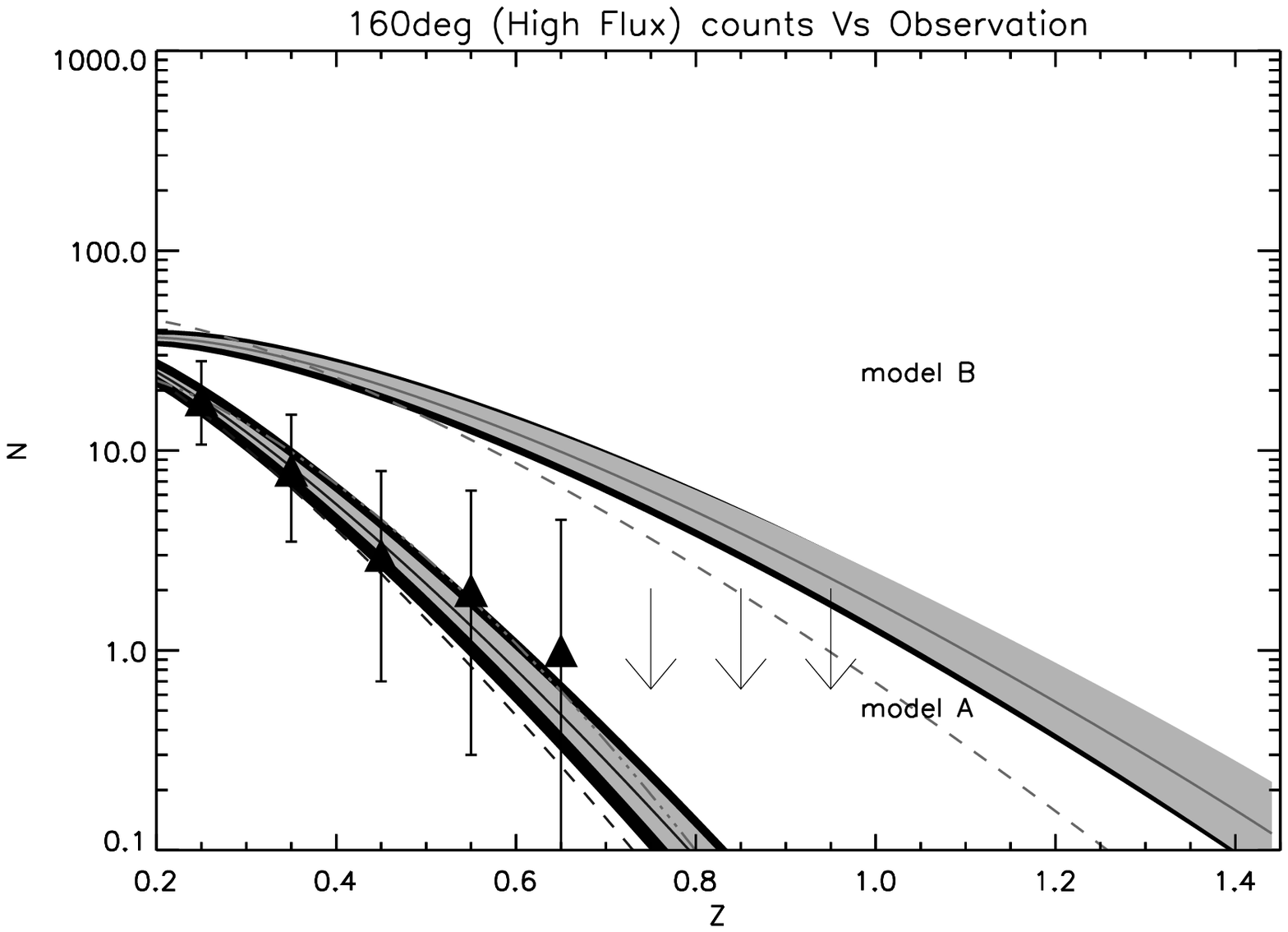} \\
\includegraphics[angle=0,totalheight=4.cm,width=5.cm]{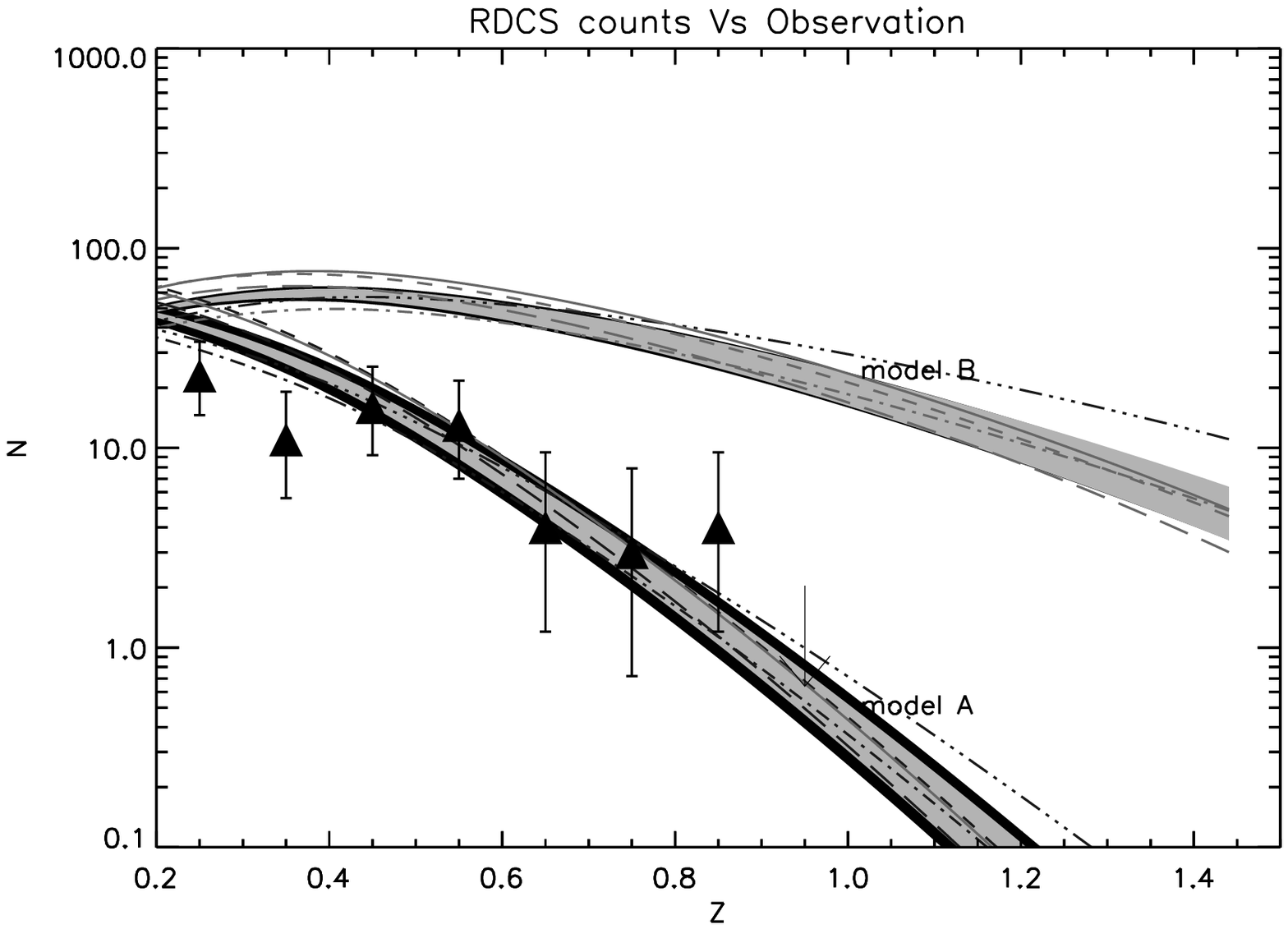}&
\includegraphics[angle=0,totalheight=4.cm,width=5.cm]{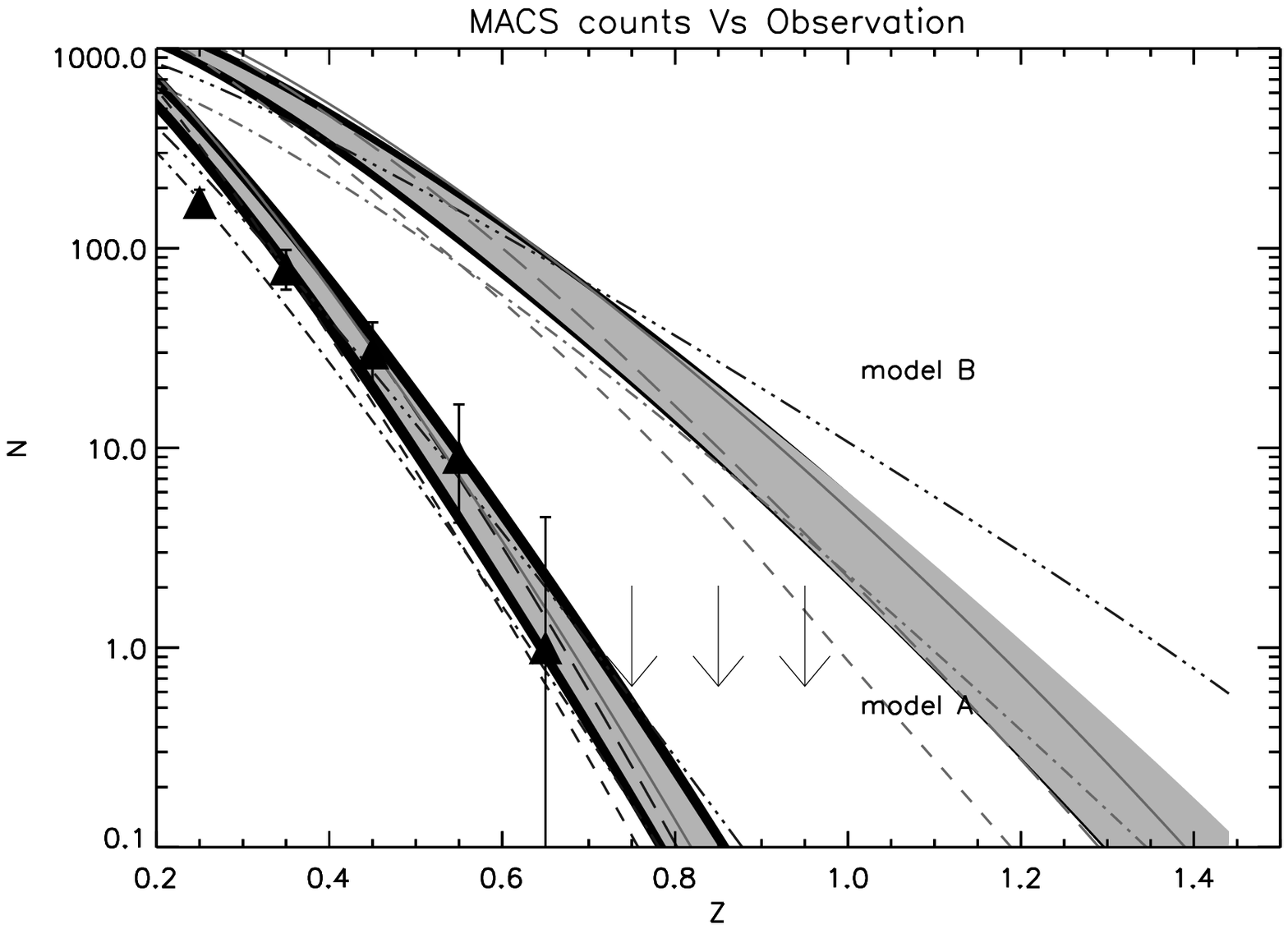} \\
\end{tabular}
\caption{\label{xmm1}  Theoretical  number counts in bins of redshift ($\Delta z=0.1$) for the  different surveys:
RDCS, EMSS,  MACS and 160deg$^2$. Observed numbers 
are triangles with 95\% confidence 
interval on the density assuming poissonian statistics (arrows are 95\% upper limits).
 For the  160deg$^2$ we show here only the brightest part ( $f_x > 2 \, 10^{-13}$ erg/s/cm$^2$). The upper curves are the predictions in the concordance 
model.The continuous lines correspond to $T_{15} = 4.$  while the dashed lines are for $T_{15} = 6.$ The grey area show our estimate on the uncertainty in the evolution of the L-T relation. The two last plots show the systematics effect: counts using  the Press and Schechter mass function (solid line) and changing the local L-T (slope and normalization dashed lines) from $0.04T^{3}$ to $0.08T^{3}$ and $L \propto T^{2.7}$ to  $L \propto T^{3.3}$.  }
\end{center}
\end{figure}

\begin{small}
\begin{table}[!t]
\begin{tabular}{lllllc} \hline

 T$_{15}$   & $\Omega_M$ & $\sigma_8$ & $\Gamma$ & Cosmological model  \\
 (keV)      &            &            &          &    and ingredients \\ \hline
  4         &  0.3       &    1.   &   0.2      & B: Low $\Omega_M$+BN98+SMT \\
 6.5      &  0.3       &    0.72   &   0.2     &  B: Low $\Omega_M$+M98+SMT\\
  4         &  1.      &    0.55   &   0.12     &  A: best model+BN98+SMT\\
 6.5      &  0.85       &    0.45   &   0.1     &  A: best model+M98+SMT\\ \hline
\end{tabular}
\caption{Models and parameters used in the number counts calculations}
\label{tab2}
\end{table}                                                                    \end{small}                       




\begin{chapthebibliography}{1}
\bibitem[\protect\astroncite{omegaP}{2000}]{OP}
Bartlett, J.  et al., 2001, proceedings of the XXI rencontre de Moriond, 
astro-ph/0106098

\bibitem[\protect\astroncite{Blanchard2}{2000}]{BSBL}
Blanchard, A., Sadat, R.,   Bartlett, J. \& Le Dour, M. 2000, A\&A 362, 809





\bibitem[\protect\astroncite{Henry}{1997}]{}
Henry, J.P.  1997, ApJ 489, L1

\bibitem[\protect\astroncite{Lumb}{2003}]{OP1}
Lumb, D. et al., 2003, A\&A, in preparation

 


\bibitem[\protect\astroncite{{Sadat}}{1996}]{}
Sadat, R., Blanchard, A. \& Oukbir, J. 1998, A\&A 329, 21

\bibitem[\protect\astroncite{{Vauclair} et al.}{2003}]{SV3}
{Vauclair}, S. C. et al., 2003, submitted  

\bibitem[\protect\astroncite{{Vikhlinin} et al.}{2002}]{V02}
{Vikhlinin}, A. et al., 2002, ApJL, 578, 107  

\end{chapthebibliography}
\end{document}